\documentclass[aps, prd, showpacs, nofootinbib, floatfix, superscriptaddress, onecolumn]{revtex4-2}
\usepackage{amsmath,amssymb,amsfonts}
\usepackage{float}
\usepackage{graphicx}
\usepackage{xcolor}
\usepackage{tensor}
\usepackage{lineno}
\usepackage[colorlinks,linkcolor=blue,citecolor=blue,urlcolor=blue]{hyperref}
\usepackage{appendix}
\graphicspath{{./Fig/}}
\begin{document}

\title{Hadron momentum spectra from analytical solutions of relativistic hydrodynamics}

%\title{Analytical solutions of relativistic hydrodynamics and hadron momentum spectra}

%\title{Transverse momentum and rapidity spectra from analytical hydrodynamic solutions}

\date{ \today }

\author{Mahammad Sabir Ali}
\affiliation{School of Physical Sciences, National Institute of Science Education and Research, An OCC of Homi Bhabha National Institute, Jatni-752050, India}

\author{Deeptak Biswas}
\affiliation{School of Physical Sciences, National Institute of Science Education and Research, An OCC of Homi Bhabha National Institute, Jatni-752050, India}
 
\author{Amaresh Jaiswal}
\affiliation{School of Physical Sciences, National Institute of Science Education and Research, An OCC of Homi Bhabha National Institute, Jatni-752050, India}
\affiliation{Institute of Theoretical Physics, Jagiellonian University, ul. St. \L ojasiewicza 11, 30-348 Krakow, Poland}

\author{Sushant K. Singh}
\affiliation{Department of Physics \& Astronomy, University of Florence, Via G. Sansone 1, 50019 Sesto Fiorentino, Florence, Italy}
\affiliation{Variable Energy Cyclotron Centre, 1/AF, Bidhan Nagar , Kolkata 700064, India}

%%%%%%%%%%%%%%%%%%%%%%%%%%%%%%%%%%%%%%%%%%%%%%%%%%%%%%

\begin{abstract}  

We present analytical solution of relativistic hydrodynamics for a system having cylindrical symmetry with boost-invariant longitudinal expansion and Hubble-like transverse expansion. We also consider analytical solution for Hubble-like spherically expanding system. For these two cases, we calculate analytical expression for transverse momentum spectra of hadrons, at constant temperature freeze-out hypersurface using Cooper-Frye prescription. We compare our results for transverse momentum spectra with experimental results from Large Hadron Collider and CERN SPS where one expects cylindrical and spherical geometry of the fireball, respectively. In the case of low-energy collisions with spherical geometry, we calculate rapidity spectra and compare with the results from CERN SPS.

\end{abstract}

%%%%%%%%%%%%%%%%%%%%%%%%%%%%%%%%%%%%%%%%%%%%%%%%%%%%%%

\maketitle

%\linenumbers

%%%%%%%%%%%%%%%%%%%%%%%%%%%%%%%%%%%%%%%%%%%%%%%%%%%%%%

\section{Introduction}

The earliest hydrodynamic approach for describing nucleus-nucleus collisions dates back to Landau's work in 1953, where he focused on longitudinal expansion along the collision axis~\cite{Landau:1953gs}. Landau's non-dissipative expansion model provided an approximate analytical solution to ideal hydrodynamic equations, known as Landau hydrodynamics, resulting in Gaussian-like rapidity distributions of produced particles \cite{Landau:1953gs, Wong:2008ex}. Subsequent developments in hydrodynamics included the Hwa-Bjorken framework, proposing a boost-invariant scenario with a plateau-like rapidity distribution \cite{Hwa:1974gn, Bjorken:1982qr}. However, the applicability of boost-invariance symmetry was limited to the mid-rapidity region in ultra-relativistic heavy-ion collisions. In contrast, the transverse-momentum ($p_T$) integrated yield across the entire rapidity region demonstrated an overall better agreement with the Gaussian structure suggested by Landau \cite{Murray:2004gh, Bearden:2004yx, Murray:2007cy, Steinberg:2004wx, Steinberg:2007iv, Wong:2008ex, Biswas:2019wtp}. Since Landau and Hwa-Bjorken solutions, several different forms of analytical solutions were proposed motivated by anticipated collective behaviour in relativistic heavy ion collisions~\cite{Akkelin:2000ex, Csorgo:2003rt, Csorgo:2006ax, Lin:2009kv, Gubser:2010ze, Csorgo:2013ksa, Bantilan:2018vjv}. Analytical solutions of relativistic hydrodynamics in simple symmetric cases are important because they serve as benchmarks for testing more realistic hydrodynamic simulation codes. In the last two decades, relativistic hydrodynamic simulations have been successfully applied to model the space-time evolution of the hot and dense matter produced at the Relativistic Heavy Ion Collider (RHIC) at BNL and the Large Hadron Collider (LHC) at CERN \cite{Hirano:2002ds, Kolb:2003dz, Romatschke:2007mq, Song:2007ux, Luzum:2008cw, Luzum:2009sb, Song:2010mg, Schenke:2010rr, Luzum:2010ag, Schenke:2011tv, Gale:2012rq, Heinz:2013th, Gale:2013da, Bhalerao:2015iya, Jaiswal:2016hex}. 

Majority of studies in hydrodynamic modelling of the evolution of hot and dense QCD matter, formed in relativistic heavy-ion collisions, rely on numerical simulations. The hydrodynamic evolution equations, comprising of energy-momentum and current conservation equations, are solved numerically with appropriate initial conditions. The conversion from hydrodynamic fields to the particle degree of freedom is achieved via the Cooper-Frye freezeout prescription~\cite{Cooper:1974mv}, which leads to the momentum distribution of observed hadrons. Semi-analytical framework of hydrodynamics-inspired phenomenological models provides another approach to study hadron momentum distribution~\cite{Siemens:1978pb, Lee:1988rd, Lee:1990sk, Schnedermann:1993ws, Florkowski:2004tn, Florkowski:2005nh}. Due to their simplicity, these phenomenological models have been employed extensively to analyze the momentum spectra of produced hadrons, offering insights into the properties of the strongly interacting matter at kinetic freeze-out~\cite{Retiere:2003kf, Csorgo:1995bi, Csorgo:1999sj, Dobler:1999ju, Huovinen:2001cy, STAR:2001ksn, Csanad:2003qa, Teaney:2003kp, Cramer:2004ih, Miller:2005ji, Renk:2004yv, Renk:2004cj, Broniowski:2001we, Broniowski:2001uk, Broniowski:2002nf, Ghosh:2014eqa, Jaiswal:2015saa, Rode:2018hlj, Rode:2023vjl}. These phenomenological models, like the Blast-wave model, primarily rely on the Cooper-Frye freezeout prescription with ad-hoc parameterization of fluid variables at the freeze-out hypersurface. While the parameterization of the fluid variables are inspired by hydrodynamic solutions, a direct correspondence between them has not yet been established. Therefore it would be highly beneficial to have a simple and fully analytical hydrodynamic framework for computation of hadron momentum distribution. 

In this work, we obtain an analytical solution for relativistic hydrodynamics in a system characterized by cylindrical symmetry with boost-invariant longitudinal expansion and Hubble-like transverse expansion. Additionally, we explore an analytical solution for a Hubble-like spherically expanding system. These straightforward profiles allow us to derive a simple analytical solution for the temperature evolution relative to the proper time, facilitating a connection with the freeze-out hypersurface. For both cylindrical and spherical flow profiles, we compute the analytical expressions for the transverse momentum spectra of hadrons on a constant temperature freeze-out hypersurface, utilizing the Cooper-Frye prescription. We then compare these results for transverse momentum spectra with experimental findings from the Large Hadron Collider (LHC) and CERN SPS, where the fireball is expected to exhibit cylindrical and spherical geometries, respectively. In the case of low energy collisions with spherical expansion, we further calculate rapidity spectra and compare them with results from CERN SPS. The paper is organized as follows, in Sect.~\ref{sec:sol_hydro} we provide the derivation of the analytical hydrodynamic solutions for spherical and cylindrical geometries. Subsequently, in Sect.~\ref{sec:FO_BW} we formulate the hadron momentum spectra using Cooper-Frye prescription for spherical and cylindrical freezeout hypersurface. Further, in Sect.~\ref{sec:results}, we apply the derived expressions for hadron momentum distribution to compare with experimental results for hadron transverse momentum spectrum and rapidity spectrum. Finally in Sect.~\ref{sec:summary}, we provide the summary and conclusion of the present work. Throughout the text we employ natural units where $ \hbar = c = k_B =1$ and use $(+,-,-,-)$ for the metric signature. We use dot to denote scalar products of both three- and four-vectors, i.e., $a \cdot b = a^0 b^0 - \vec{a} \cdot \vec{b}$.

%%%%%%%%%%%%%%%%%%%%%%%%%%%%%%%%%%%%%%%%%%%%%%%%%%%%%%

\section{Analytic solution of hydrodynamics}\label{sec:sol_hydro}

Hydrodynamic evolution of the hot and dense QCD matter is governed by the conservation equations for particle charge current as well as energy and momentum conservation. Assuming a system with no conserved charges, the hydrodynamical equations are 
\begin{equation}\label{eqn:hydro}
    \partial_\mu T^{\mu\nu}(x)=0,
\end{equation}
where $T^{\mu\nu}(x)$ is the local energy-momentum tensor. In the following, we consider non-dissipative evolution of the system where the energy-momentum tensor takes the form
\begin{equation}\label{eqn:eng_mom_tens}
T^{\mu\nu}=(\varepsilon+p)u^\mu u^\nu  - pg^{\mu\nu},
\end{equation}
where $\varepsilon$ and $p$ denote the energy density and pressure of the system, respectively. The local four-velocity is denoted by $u^\mu (x) = \gamma(x) (1,\vec{v}(x))$, with $\gamma(x)=1/\sqrt{1-\vec{v}^2(x)}$ being the local Lorentz factor. We use the Minkowski metric for our calculations, given by $g^{\mu\nu}={\rm diag}(1,-1,-1,-1)$. 

Using the thermodynamic expressions $Ts = \varepsilon +p$ and $dp = s\,dT$, we can express Eq.~\eqref{eqn:hydro} in terms of temperature $T$ as 
\begin{equation}\label{eqn:therm_hyd}
    u^\mu \partial_\mu (Tu^\nu) - \partial^\nu T = 0,
\end{equation}
where $s$ is the entropy density. The above equation can be re-written in the following form
\begin{equation}\label{eqn:curlT}
    \frac{\partial }{\partial t}(T\gamma \vec{v}) + \vec{\nabla} (T\gamma) = \vec{v}\times \vec{\nabla} \times (T\gamma \vec{v}).
\end{equation}
For non-dissipative fluid, the entropy conservation implies
\begin{equation}\label{eqn:entr_cons}
    \partial_\mu (s u ^\mu) = 0,
\end{equation}
which can be re-written in the following form
\begin{equation}\label{eqn:entropyconsrv}
    \frac{\partial}{\partial t}(s\gamma) + \vec{\nabla} \cdot (s\gamma \vec{v}) = 0.
\end{equation}
In the following, we will seek solutions to Eqs.~\eqref{eqn:curlT} and \eqref{eqn:entropyconsrv} for a hydrodynamically expanding system with spherical and cylindrical symmetry.

%%%%%%%%%%%%%%%%%%%%%%%%

\subsection{Spherical solution}

In the case of radially expanding system, it is convenient to work in the $(t,r,\theta,\phi)$ co-ordinates. Due to spherical symmetry, the hydrodynamic variables depend only on time $t$ and radial distance $r$ from the origin. In this case, the fluid velocity vector is given by $\vec{v}=(v_r,0,0)$. For such a system, Eqs.~\eqref{eqn:curlT}~and~\eqref{eqn:entropyconsrv} become
\begin{align}  
    \frac{\partial }{\partial t}(T\gamma v_r) + \frac{\partial }{\partial r}(T\gamma ) &= 0 \label{eqn:ssvr} \\
     \frac{\partial }{\partial t}(s\gamma ) + \frac{1}{r^2}\frac{\partial }{\partial r}(r^2 s\gamma v_r) &= 0. \label{eqn:ss2}
\end{align}
Above equations admit scaling type analytical solution with the fluid flow velocity of the form $v_r = r/t$~\cite{Bondorf:1978kz, Baym:1983amj}. Assuming this flow profile, Eqs.~\eqref{eqn:ssvr} and \eqref{eqn:ss2}, become 
\begin{align} 
    r\frac{\partial T}{\partial t} + t\frac{\partial T}{\partial r} &= 0 \label{eqn:ssvr1}\\
    t\frac{\partial s}{\partial t} + r\frac{\partial s}{\partial r}+ 3s &= 0. \label{eqn:ssvr2}
\end{align}
We now require the equation of state relating energy density and pressure. We assume an equation of state of the form $p=c_s^2\,\varepsilon$, with the speed of sound $c_s$ kept constant. 

By employing thermodynamic relations, we reformulate the equation of state to relate entropy density and temperature as $s\propto T^{1/c_s^2}$. Substituting this expression for $s$ in Eq.~\eqref{eqn:ssvr2}, we obtain
\begin{equation}\label{eqn:ssvr3}
    t\frac{\partial T}{\partial t} + r\frac{\partial T}{\partial r}+ 3\, c_s^2\, T = 0.
\end{equation}
The above equation can be solved using the method of characteristics. The equation for obtaining the characteristic curves is given by
\begin{equation}\label{eqn:char_sph}
    \frac{dt}{t} = \frac{dr}{r} = -\frac{dT}{3\,c_s^2\, T}.
\end{equation}
Solving the above equation, we obtain
\begin{equation}\label{eqn:char_sph_T}
    T(r,t) \sim t^{-3c_s^2}\, f\!\left(\frac{r}{t}\right),
\end{equation}
The above equation implies that any functional form for $f(r/t)$ is a solution of Eq.~\eqref{eqn:ssvr3}. This functional form can be fixed by employing Eq.~\eqref{eqn:ssvr1}. Substituting the expression for $T(r,t)$ from Eq.~\eqref{eqn:char_sph_T} in Eq.~\eqref{eqn:ssvr1}, we get
\begin{equation}\label{eqn:char_sph_fn}
    f'(w)(1-w^2) - 3c_s^2 w f(w) = 0,
\end{equation}
where we have defined $w\equiv r/t$. The above equation can be easily integrated to obtain
\begin{equation}\label{eqn:sph_char_fn}
    f(w) \sim \frac{1}{(1-w^2)^{3c_s^2/2}}.
\end{equation}
Using the above equation in the expression for $T(r,t)$ and $s(r,t)$, we obtain the solution to Eqs.~\eqref{eqn:ssvr} and \eqref{eqn:ss2} as
\begin{align}
    T(r,t) &= T_0\left[\frac{t_0^2-r_0^2}{t^2-r^2} \right]^{3c_s^2/2}, \label{eqn:sph_sol_T} \\
    s(r,t) &= s_0\left[\frac{t_0^2-r_0^2}{t^2-r^2} \right]^{3/2}, \label{eqn:sph_sol_s} \\
    v_r &= \frac{r}{t}, \label{eqn:sph_sol_vr}
\end{align}
where $T_0$ and $s_0$ are constants denoting the initial temperature and entropy density, respectively, at position $r_0$ and initial time $t_0$. We note that the scaling solution, given in Eqs.~\eqref{eqn:sph_sol_T}-\eqref{eqn:sph_sol_vr}, is in agreement with the family of solutions obtained in Refs.~\cite{Csorgo:2003rt, Lin:2009kv}.

%%%%%%%%%%%%%%%%%%%%%%%%

\subsection{Cylindrical solution}

In the case of a longitudinally and transversely expanding system, it is convenient to work in the $(t,\rho,\phi,z)$ coordinates. Due to cylindrical symmetry, the hydrodynamic variables depend only on time $t$, radial distance $\rho$ from the origin in the transverse plane, and the longitudinal distance $z$. In this case, the fluid velocity vector is given by $\vec{v}=(v_\rho,0,v_z)$. For such a system, Eqs.~\eqref{eqn:curlT} and \eqref{eqn:entropyconsrv} leads to
\begin{align}
    \frac{\partial }{\partial t}(T\gamma v_\rho) + \frac{\partial }{\partial \rho}(T\gamma ) - v_z \left[ \frac{\partial }{\partial \rho}(T\gamma v_z) - \frac{\partial }{\partial z}(T\gamma v_\rho)  \right] &= 0 \label{eqn:cs1} \\
    \frac{\partial }{\partial t}(T\gamma v_z) + \frac{\partial }{\partial z}(T\gamma ) - v_\rho \left[ \frac{\partial }{\partial z}(T\gamma v_\rho) - \frac{\partial }{\partial \rho}(T\gamma v_z)  \right] &= 0 \label{eqn:cs2} \\
    \frac{\partial }{\partial t}(s\gamma ) + \frac{1}{\rho}\frac{\partial }{\partial \rho}(\rho s\gamma v_\rho) +\frac{\partial }{\partial z}(s\gamma v_z) &= 0. \label{eqn:cs3}
\end{align}
In the following, we consider boost-invariance for the longitudinal expansion, i.e., $v_z = \frac{z}{t}$. Bjorken's notion of boost-invariance implies that the expansion geometry of the fireball is identical in all longitudinally boosted frames. Therefore, considering the $z=0$ slice, Eqs.~\eqref{eqn:cs1}-\eqref{eqn:cs3} reduces to
\begin{align}
    \frac{\partial }{\partial t}(T\gamma v_\rho) + \frac{\partial }{\partial \rho}(T\gamma ) &= 0, \label{eqn:bi1} \\
    \frac{\partial }{\partial t}(s\gamma ) + \frac{\partial }{\partial \rho}(s\gamma v_\rho) + s\gamma \left( \frac{v_\rho}{\rho}+\frac{1}{t}\right) &= 0. \label{eqn:bi2} 
\end{align}
As done in the previous case of spherical expansion, we again assume an equation of state of the form $p=c_s^2\,\varepsilon$ with constant $c_s^2$. 

In order to find the solution of Eqs.~\eqref{eqn:bi1}-\eqref{eqn:bi2}, we introduce the potential $\Phi (T)$ and define $a_\pm$ as~\cite{Baym:1983amj}
\begin{align}
    d\Phi &= \frac{d \ln T}{c_s} = c_s \ d\ln s, \label{eqn:dPhi} \\
    a_\pm &= e ^{\Phi \pm \alpha}, \label{eqn:apm}
\end{align}
where $\alpha$ denotes the transverse rapidity, i.e., $\alpha = \tanh^{-1}(v_\rho)$. The above equations can be inverted to write the hydrodynamic variables in terms of these new variables $a_\pm$ as
\begin{align}
    \Phi (T) &= \frac{1}{2}\ln a_+ a_-, \label{eqn:Phi} \\
    v_\rho &= \frac{a_+-a_-}{a_++a_-}. \label{eqn:vrho}
\end{align}
In terms of the new variables $a_\pm$, Eqs.~\eqref{eqn:bi1} and \eqref{eqn:bi2} can be rewritten as~\cite{Baym:1983amj}
\begin{align}
    \frac{\partial a_+}{\partial t}+\frac{v_\rho+c_s}{1+v_\rho\, c_s}\,\frac{\partial a_+}{\partial \rho}+\frac{c_s a_+}{1+v_\rho\, c_s} \left( \frac{v_\rho}{\rho}+\frac{1}{t} \right)&=0, \label{eqn:a+} \\
    \frac{\partial a_-}{\partial t}+\frac{v_\rho-c_s}{1-v_\rho\,c_s}\,\frac{\partial a_-}{\partial \rho}+\frac{c_s a_-}{1-v_\rho\, c_s}\left( \frac{v_\rho}{\rho}+\frac{1}{t} \right) &=0. \label{eqn:a-}
\end{align}
We intend to find a self-similar solution for the above set of equations.

We start with a simple scaling ansatz for the transverse component of the fluid velocity, $v_\rho = \rho/t$. Using this ansatz in Eq.~\eqref{eqn:vrho}, the variables $a_\pm$ become
\begin{align}
    a_+(\rho,t) &= A(\rho,t)\left( 1+\frac{\rho}{t} \right), \label{eqn:ap} \\
    a_-(\rho,t) &= A(\rho,t)\left( 1-\frac{\rho}{t} \right), \label{eqn:am}
\end{align}
where $A(\rho,t)$ is a constant of proportionality. Substituting the above expressions in Eqs.~\eqref{eqn:a+} and \eqref{eqn:a-}, we get
\begin{align}
    t(t+ \rho c_s)\frac{\partial A}{\partial t}  + t(\rho+t c_s) \frac{\partial A}{\partial \rho} + A c_s (3t - \rho) &= 0 \label{eqn:dA1} \\
    t(t- \rho c_s)\frac{\partial A}{\partial t}  + t(\rho-t c_s) \frac{\partial A}{\partial \rho} + Ac_s (3t + \rho) &= 0 \label{eqn:dA2}
\end{align}
Assumption of the form of the scaling solution $v_\rho = \rho/t$, along with Eqs.~\eqref{eqn:a+} and \eqref{eqn:a-}, puts constraints on $A(\rho,t)$. The form of $A(\rho,t)$ must be such that it satisfies the above differential equations. By adding and subtracting Eqs.~\eqref{eqn:dA1} and \eqref{eqn:dA2}, we obtain a simpler set of equations given by 
\begin{align}
    \label{eqn:adda+a-} \frac{\partial A}{\partial t}  + \frac{\rho}{t} \frac{\partial A}{\partial \rho} + \frac{3 A c_s}{t}=0, \\
    \label{eqn:suba+a-} \frac{\partial A}{\partial t}  + \frac{t}{\rho} \frac{\partial A}{\partial \rho} - \frac{A}{t}  = 0.
\end{align}
The above equations can be solved using the method of characteristics.

Let us first consider the partial differential equation in Eq.~\eqref{eqn:adda+a-}. The characteristic curves are obtained by solving the equations
\begin{equation}\label{eqn:char_cyl}
    \frac{dt}{t} = \frac{d\rho}{\rho} = -\frac{dA}{3\,c_s A}.
\end{equation}
It is interesting to note that the above equations for characteristic curve are similar in structure to Eq.~\eqref{eqn:char_sph}, suggesting that they belong to a broader family of solutions~\cite{Csorgo:2003rt}. The solution of the above equations is given by
\begin{equation}\label{eqn:Ag}
    A(\rho,t) \sim t^{-3c_s}\, g\!\left(\frac{\rho}{t}\right),
\end{equation}
where $g(\rho/t)$ is an arbitrary function. The form of $g(\rho/t)$ is obtained by substituting Eq.~\eqref{eqn:Ag} in Eq.~\eqref{eqn:suba+a-}. Defining $w \equiv \rho/t$, we get
\begin{equation}
    \left(\frac{1}{w}-w\right) g'(w)  -(1+3c_s) g(w) = 0.
\end{equation}
The above equation can be easily integrated to obtain $g(w)$ which leads to 
\begin{equation}
    A(\rho,t) \sim \left(1-\frac{\rho^2}{t^2}\right)^{-(1+3c_s)/2} t^{-3c_s}.
\end{equation}
Using this result for $A(\rho,t)$ in Eqs.~\eqref{eqn:Phi}, \eqref{eqn:ap} and \eqref{eqn:am}, 
we get
\begin{equation} \label{eqn:Phi_fin}
    \Phi (T) \sim - 3\,c_s \ln(t^2 -\rho^2).
\end{equation}
Substituting the above expression in Eq.~\eqref{eqn:dPhi}, we obtain the solutions for temperature and entropy density as,
\begin{align}
    T(\rho,t) &= T_0\left[\frac{t_0^2-\rho_0^2}{t^2-\rho^2}\right]^{3c_s^2/2}, \label{eqn:T_cyl}\\
    s(\rho,t) &= s_0\left[\frac{t_0^2-\rho_0^2}{t^2-\rho^2}\right]^{3/2}. \label{eqn:s_cyl}
\end{align}
where $T_0$ and $s_0$ are constants denoting the initial temperature and entropy density, respectively, at position $\rho_0$ and initial time $t_0$. Note that the solution given in Eqs.~\eqref{eqn:T_cyl} and \eqref{eqn:s_cyl} has been derived for the $z=0$ slice. It is then straightforward to show that 
\begin{align}
    T(\rho,z,t) &= T_0\left[\frac{t_0^2-\rho_0^2-z_0^2}{t^2-\rho^2-z^2}\right]^{3c_s^2/2},  \label{eqn:T_cyl_fin}\\
    s(\rho,z,t) &= s_0\left[\frac{t_0^2-\rho_0^2-z_0^2}{t^2-\rho^2-z^2}\right]^{3/2}, \label{eqn:s_cyl_fin}\\
    v_\rho (\rho,z,t) &= \frac{\rho}{t}, \label{eqn:vrho_fin}\\
    v_z (\rho,z,t) &= \frac{z}{t}. \label{eqn:vz_fin}
\end{align}
form a solution of Eqs.~\eqref{eqn:cs1}-\eqref{eqn:cs3} for all $z$. We note that the scaling solution, given in Eqs.~\eqref{eqn:T_cyl_fin}-\eqref{eqn:vz_fin}, is in agreement with the family of solutions obtained in Refs.~\cite{Csorgo:2003rt, Lin:2009kv}.

We would like to reiterate that the solutions in Ref.~\cite{Baym:1983amj} are obtained numerically without assuming a specific velocity profile, by solving Eq.~\eqref{eqn:vrho}. While these non-trivial profiles provide a more detailed understanding, we believe that a simplified analytical solution offers a valuable baseline for studying the evolution, complementing more complex numerical approaches. We note that the analytical solutions obtained in Eqs.~\eqref{eqn:sph_sol_T} and \eqref{eqn:T_cyl_fin} are scaling in nature. Defining $\tau_3=\sqrt{t^2-r^2} =\sqrt{t^2-\rho^2-z^2}$, we find that the initial conditions corresponds to $T=T_0$, $s=s_0$ at initial proper time $\tau_3(t_0,r_0)$ and $\tau_3(t_0,\rho_0,z_0)$ for the spherical and cylindrical case, respectively. Additionally, we emphasize that our findings are in line those from Ref.~\cite{Baym:1983amj}, which indicate that a scaling solution is valid when both $r$ and $t$ are large compared to the initial size of the system.

%%%%%%%%%%%%%%%%%%%%%%%%%%%%%%%%%%%%%%%%%%%%%%%%%%%%%%
\section{Freeze-out and momentum spectra}
\label{sec:FO_BW}
%%%%%%%%%%%%%%%%%%%%%%%%%%%%%%%%%%%%%%%%%%%%%%%%%%%%%%

The hadron momentum spectra can be derived by applying the Cooper-Frye prescription for particle production at the freezeout hypersurface
\begin{equation}\label{eqn:CF}
    E\frac{dN}{d^3p} = \frac{g}{(2\pi)^3}\int p_\mu d\Sigma^\mu f(x,p),
\end{equation}
where $g$ is the particle degeneracy factor\,\footnote{Not to be confused with $g(w)$ introduced in the previous section.}, $E$ is the relativistic particle energy, $d\Sigma_\mu$ is the freeze-out hyper-surface, $p^\mu$ is the particle momenta and $f(x,p)$ is the phase-space distribution function of the particles. In this work, we consider the form of the equilibrium distribution function to be semi-classical generalization of Maxwell-J\"uttner distribution,
\begin{equation}\label{eq:dist_fn_eq}
    f_{\rm eq}(x,p) = \frac{1}{ \exp [\beta(u\cdot p)] + \epsilon},
\end{equation}
where $\epsilon=-1$ for Bose-Einstein statistics, $\epsilon=+1$ for Fermi-Dirac statistics and $\epsilon=0$ for classical Maxwell-Boltzmann statistics. In the above equation, we have defined $\beta\equiv1/T$ as the inverse temperature. For ease of calculation, we rewrite Eq.~\eqref{eq:dist_fn_eq} in the form 
\begin{equation}\label{eq:dist_fn_ser}
    f_{\rm eq}(x,p) = \sum_{n=1}^{\infty}\epsilon_n \exp [-n\beta(u\cdot p) ],
\end{equation}
where, $\epsilon_n\equiv(-\epsilon)^{n-1}$ with $\epsilon_1=1$ for classical Maxwell-Boltzmann statistics.

To proceed further, we need to specify the freeze-out hypersurface for the case of spherical and cylindrical expansion. Defining $(3+1)$-dimensional proper time $\tau_3 = \sqrt{t^2-r^2} = \sqrt{t^2-\rho^2-z^2}$, we see from Eqs.~\eqref{eqn:sph_sol_T} and \eqref{eqn:T_cyl_fin} that analytical solution for temperature, in case of both spherical and cylindrical expansion, leads to
\begin{equation}\label{eqn:T_tau3}
    T(\tau_3) = T_0\left( \frac{\tau_{30}}{\tau_3} \right)^{3c_s^2}.
\end{equation}
From here, we can easily conclude that $T_0$ is the initial temperature at initial proper time $\tau_{30}$. The above equation also suggests that the scaling nature of the solution determines the initial temperature profile of the fireball. 

Kinetic freeze-out in heavy-ion collision is considered to occur when the constituent hadrons stop interacting and free stream to the detectors. In hydrodynamic simulations, this is imposed when the temperature reaches the kinetic freeze-out temperature due to the cooling caused by the expansion of the fireball. In the present work, we also assume that the freeze-out hypersurface is specified at a constant temperature $T_f$. From Eq.~\eqref{eqn:T_tau3}, we see that the $T=T_f$ hypersurface uniquely corresponds to a constant proper time hypersurface $\tau_3=\tau_{3f}$. In the following, we calculate this freeze-out hypersurface and the corresponding hadron momentum spectra for the two cases of spherical and cylindrical expansion.

%%%%%%%%%%%%%%%%%%%%%%%%

\subsection{Spherical geometry}

Points on a spherically symmetric hypersurface can be parametrized as~\cite{Florkowski:2010zz}
\begin{equation}\label{eqn:sph_pnts}
    x^\mu = \left[ t(\zeta),r(\zeta)\sin \theta \cos \phi,r(\zeta)\sin \theta \sin \phi,r(\zeta)\cos \theta \right],
\end{equation}
where $r,\,\theta,\,\phi$ are usual spherical polar coordinate variables. Here $\zeta$ is the parameter which defines the spherical hypersurface with $0<\zeta <1$ such that $r(0)=0$ and $r(1)=R$, with $R$ being the radius of the fireball at freezeout. The surface element of the hypersurface can be obtained as
\begin{equation}\label{eqn:sph_hyp}
    d\Sigma^\mu = \left(\frac{dr}{d\zeta}, \frac{dt}{d\zeta}\sin \theta \cos \phi,\frac{dt}{d\zeta}\sin \theta \sin \phi,\frac{dt}{d\zeta}\cos \theta \right) r^2 \sin\theta \ d\zeta \ d\theta\ d\phi.
\end{equation}
For radially expanding fireball, the spherically symmetric flow velocity profile can be written as
\begin{equation}\label{eqn:sph_umu_param}
    u^\mu = \gamma (\zeta) \left[ 1, v_r(\zeta)\sin \theta \cos \phi,v_r(\zeta)\sin \theta \sin \phi,v_r(\zeta)\cos \theta \right],
\end{equation}
Where we recall that we have Hubble-like solution for the fluid flow velocity, $v_r(\zeta)=\frac{r(\zeta)}{t}$.

Using spherical polar co-ordinate variables in momentum space, the four-momentum of a hadron, $p^\mu=(E,\,p^x,\,p^y,\,p^z)$, can be written as
\begin{equation}\label{eqn:sph_pmu}
    p^\mu = (E, p\sin \theta_p \cos\phi_p, p\sin \theta_p\sin \phi_p,p\cos\theta_p ),
\end{equation}
where $\theta_p$ and $\phi_p$ denote the polar and azimuthal angle, respectively, in the momentum space. Spherical symmetry
can be employed to restrict our considerations to $\theta_p =0$, which leads to
\begin{align}
    u \cdot p &= \gamma(\zeta) \, E - \gamma(\zeta)\, v_r(\zeta)\, p \cos \theta, \label{eqn:sph_udotp} \\
    p \cdot d\Sigma &= \left( E\,\frac{dr}{d\zeta} - p\cos\theta\, \frac{dt}{d\zeta} \right)r^2(\zeta) \sin\theta \, d\zeta \, d\theta \, d\phi. \label{eqn:sph_pdotdS}
\end{align}
Substituting the above expressions in Eq.~\eqref{eqn:CF}, along with the form of the equilibrium phase-space distribution function given in Eq.~\eqref{eq:dist_fn_ser}, we get
\begin{equation}\label{eqn:sph_mom_dist1}
    E\frac{dN}{d^3p} = \frac{g}{(2\pi)^2} \sum_{n=1}^\infty \epsilon_n\,  \int_0^1 d\zeta \ r^2 \int_0^\pi d\theta \, \sin\theta \left( E\, \frac{dr}{d\zeta} - p\cos\theta\, \frac{dt}{d\zeta}\right) e^{-n\beta (u\cdot p)}.
\end{equation}
We notice that the $\theta$ integral in the above equation can also be done analytically.

After performing the $\theta$ integral in Eq.~\eqref{eqn:sph_mom_dist1}, the expression for the momentum distribution is given by
\begin{equation}\label{eqn:sph_mom_dist2}
    E \frac{dN}{d^3p} = \frac{g}{2\pi^2} \sum_{n=1}^\infty \epsilon_n\, \int_0^1 e^{-n\beta\gamma E}\left[E \frac{dr}{d\zeta}\frac{\sinh (n\alpha)}{n\alpha}+ T\frac{dt}{d\zeta}\left(\frac{\sinh(n\alpha) - n\alpha\cosh (n\alpha)}{n^2\alpha\gamma v_r}\right)\right]r^2\, d\zeta,
\end{equation}
where $\alpha\equiv\beta \gamma v_r p$. To proceed further, we make use of the condition for constant temperature freezeout hypersurface resulting in $\tau^2_{3f}=\text{const.}=t^2 - r^2$. This choice of freezeout hypersurface leads to
\begin{equation}\label{eqn:sph_fo_cond}
    \frac{dt}{d\zeta}= \frac{r}{t} \, \frac{dr}{d\zeta}.
\end{equation}
Using the above relation, we obtain the final form of the particle momentum distribution to be
\begin{equation}\label{eqn_pT_sph}
    E\frac{dN}{d^3p} = \frac{gR^3}{2\pi^2} \sum_{n=1}^\infty \epsilon_n\, \int_0^{1} e^{-n\beta E\sqrt{\nu^2+\chi^2}/\nu}\left[E\,\frac{\sinh (na\chi)}{na\chi}+ T\nu\left(\frac{\sinh(na\chi) - na\chi\cosh (na\chi)}{n^2 a \chi \sqrt{\nu^2+\chi^2} }\right)\right]\chi^2\ d\chi,
\end{equation}
where $\chi \equiv r/R$, $\nu \equiv \tau_{3\text{f}}/R$, and $a \equiv p/(T\nu)$. At this point, we would like to emphasize that the above expression for hadron spectrum is the exact analytical hydrodynamic result for spherically expanding fireball produced in relativistic heavy-ion collisions. 

It is instructive to calculate the average radial velocity of the fireball in terms of the parameters on the freezeout hypersurface. The average radial velocity can be defined as
\begin{equation}\label{eqn:sph_avr_def}
    \langle v_r \rangle = \frac{\int v_r \sqrt{d\Sigma_\mu d\Sigma^\mu}}{\int \sqrt{d\Sigma_\mu d\Sigma^\mu}}.
\end{equation}
With $v_r=r/t$ and employing Eqs.~\eqref{eqn:sph_hyp} and \eqref{eqn:sph_fo_cond}, we obtain the final expression for the average radial velocity as
\begin{equation}\label{eqn:sph_avr}
    \langle v_r \rangle = \frac{1 + \nu^2 \log\left(\frac{\nu^2}{1+\nu^2}\right)}{\sqrt{1+\nu^2}-\nu^2\log\left[ \frac{\nu}{\sqrt{1+\nu^2}-1}\right]}.
\end{equation}
It is straightforward to verify that the maximum value of $\langle v_r \rangle$ is $1$ for $\nu=0$.

%%%%%%%%%%%%%%%%%%%%%%%%

\subsection{Cylindrical geometry}

Points on a longitudinally boost-invariant and cylindrically symmetric hypersurface can be parametrized as~\cite{Florkowski:2010zz}
\begin{equation}\label{eqn:cyl_pnts}
    x^\mu = \left[ \tau (\zeta)\cosh \eta_s, \, \rho(\zeta) \cos \phi, \, \rho(\zeta)\sin \phi, \, \tau(\zeta)\sinh \eta_s \right]
\end{equation}
where $\rho=\sqrt{x^2+y^2}$ and $\tau = \sqrt{t^2-z^2}$. Here again, $\zeta$ is the parameter which defines the cylindrical hypersurface with $0<\zeta <1$ such that $\rho(0)=0$ and $\rho(1)=R$, with $R$ being the transverse radius of the fireball at freezeout\,\footnote{We use the same notation for transverse freezeout radius as well as spherical radius at freezeout as introduced in the previous section, for reasons that will become apparent later.}. Note that space-time rapidity $\eta_s$ is independent of the parameter $\zeta$ owing to the boost-invariance symmetry considered here. The surface element of the hypersurface can be obtained as
\begin{equation}\label{eqn:cyl_hyp}
    d\Sigma^\mu = \left( \frac{d\rho}{d\zeta}\cosh\eta_s, \,\frac{d\tau}{d\zeta}\cos\phi, \, \frac{d\tau}{d\zeta}\sin\phi, \, \frac{d\rho}{d\zeta}\sinh\eta_s\right)\tau(\zeta) \rho(\zeta)d\zeta d\phi d\eta_s.
\end{equation}
Defining longitudinal fluid flow rapidity, $\eta_f$, and transverse flow rapidity, $\eta_T$ as
\begin{align}
    \eta_f &= \frac{1}{2}\log\left(\frac{1+v_z}{1-v_z}\right), \label{eqn:etaf_def} \\
    \eta_T &= \tanh^{-1}\left(\frac{v_\rho}{\sqrt{1-v_z^2}}\right), \label{eqn:etat_def}
\end{align}
the fluid four-velocity can be written as
\begin{equation}\label{eqn:cyl_umu}
    u^\mu = \left( \cosh\eta_T \cosh\eta_f, \,\sinh\eta_T \cos\phi, \, \sinh\eta_T \sin \phi, \, \cosh\eta_T \sinh\eta_f \right).
\end{equation}
In terms of the transverse momentum and rapidity variables, the four-momentum of a particle, $p^\mu=(E,\,p^x,\,p^y,\,p^z)$, is given by
\begin{equation}
    p^\mu = \left( m_T\cosh y, \, p_T\cos \phi_p, \, p_T \sin\phi_p, \, m_T\sinh y \right),
\end{equation}
where $y$ is the longitudinal rapidity of the particle and $\phi_p$ is the azimuthal angle in momentum space. Here $m_T\equiv\sqrt{m^2+p_T^2}$ is the transverse mass with $p_T$ being the transverse momentum of the particles.

The longitudinal boost-invariance for fluid velocity profile is guaranteed for $v_z = z/t$ which leads to  $\eta_f=\eta_s$. Imposing this boost-invariance condition, we have
\begin{align}
    u\cdot p &= m_T\cosh(\eta_T)\cosh(y-\eta_s)-p_T\sinh(\eta_T)\cos(\phi-\phi_p), \label{eqn:cyl_udotp} \\
    p \cdot d\Sigma &= \left[m_T\cosh(y-\eta_s)\frac{d\rho}{d\zeta} -p_T\cos (\phi-\phi_p)\frac{d\tau}{d\zeta}\right]\tau(\zeta) \, \rho(\zeta) \, d\zeta \, d\phi  \, d\eta_s .\label{eqn:cyl_pdotdS}
\end{align}
Substituting the above expressions in Eq.~\eqref{eqn:CF}, along with the form of the equilibrium phase-space distribution function given in Eq.~\eqref{eq:dist_fn_ser}, and performing the $\phi$ and $\eta_s$ integrations, we get
\begin{equation}\label{eqn:cyl_mom_spec}
    E\,\frac{dN}{d^3p} = \frac{g}{2\pi^2}\! \sum_{n=1}^\infty\! \epsilon_n \!\int_0^1 \! \left[ m_T \, I_0(n\beta p_T\sinh \eta_T) \, K_1(n\beta m_T\cosh\eta_T) \frac{d\rho}{d\zeta} - p_T \, I_1(n\beta p_T\sinh \eta_T) \, K_0(n\beta m_T\cosh\eta_T) \frac{d\tau}{d\zeta}\right]\! \rho  \tau  d\zeta,
\end{equation}
where, $I_n(x)$ and $K_n(x)$ are modified Bessel functions of first kind and second kind, respectively.

%------------------------
\begin{figure}[t]
\includegraphics[width=9.0cm]{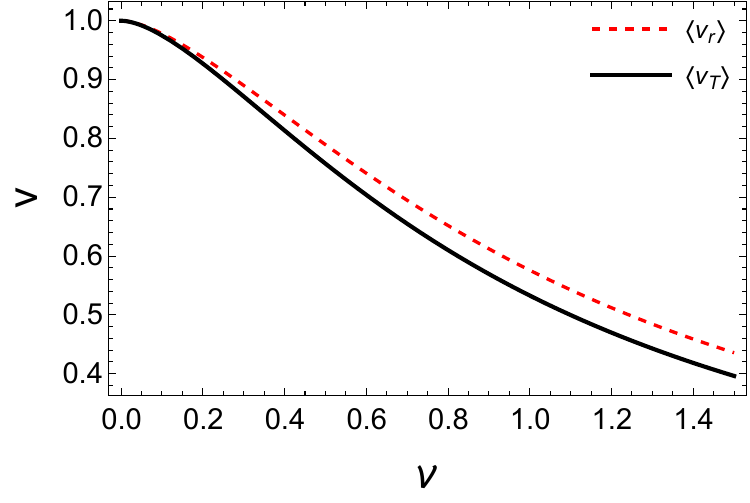}
\caption{Average radial and transverse velocity as a function of the freezeout parameter $\nu$. These average velocities are calculated using Eqs.~\eqref{eqn:sph_avr} and \eqref{eqn:cyl_avr}for spherical and cylindrical expansion geometries, respectively.}
\label{fig:avgv}
\end{figure}
%------------------------

To proceed further, we make use of the condition for constant temperature freezeout hypersurface resulting in $\tau^2_{3f}=\text{const.}=\tau^2 - \rho^2$. This choice of freezeout hypersurface leads to
\begin{equation}\label{eqn:cyl_fo_cond}
    \frac{d\tau}{d\zeta} = \frac{\rho}{\tau} \, \frac{d\rho}{d\zeta}.
\end{equation}
Considering Hubble-like flow in the transverse plane, i.e., $v_\rho = \rho/t$ leads to $\rho=\tau_{3f}\sinh \eta_T,\,\tau = \tau_{3f}\cosh \eta_T$ on the freezeout hypersurface. Using Eq.~\eqref{eqn:cyl_fo_cond} and these relations, we obtain the final form of the particle momentum distribution to be
{\small
\begin{equation}\label{eqn_pT_cyl}
    E\,\frac{dN}{d^3p} = \frac{gR^3}{2\pi^2} \sum_{n=1}^\infty \epsilon_n \!\int_0^{1}\! \left[m_T \sqrt{\nu^2+\chi^2}I_0\!\left(\!\frac{n\beta p_T \chi}{\nu}\!\right) K_1\!\left(\!n\beta m_T \sqrt{1+\frac{\chi^2}{\nu^2}}\right)\! -  \chi\, p_T I_1\!\left(\!\frac{n\beta p_T \chi}{\nu}\!\right) K_0\!\left(\!n\beta m_T \sqrt{1+\frac{\chi^2}{\nu^2}}\right)\right] \chi \, d\chi ,
\end{equation}  
}
where $\chi \equiv \rho/R$ and $\nu \equiv \tau_{3f}/R$. Comparing the above equation with Eq.~\eqref{eqn_pT_sph}, we find that the form of the overall normalization factor, appearing outside the integration on r.h.s, is identical. At this juncture, we would again like to emphasize that the above expression for the hadron spectrum is the exact analytical hydrodynamic result for a cylindrically expanding fireball produced in relativistic heavy-ion collisions.

%%%%%%%%%%%%%%%%%%%%%%%%%%%%
In the simplified model presented, the speed of sound ($c_s$) does not explicitly appear in the particle spectra. However, here we assume a constant $c_s$ throughout the evolution, which influences the temperature profile's slope, as shown in Eqs.(\ref{eqn:sph_sol_T}) and (\ref{eqn:T_cyl_fin}). The spectra are calculated at a fixed freeze-out temperature, where this temperature also implies a constant freeze-out proper time, crucial for determining the spectra. This can be understood in two steps: first, the temperature evolution is governed by the equation of state (e.o.s.), where $c_s$ plays a significant role; second, the spectra are determined at freeze-out, where the temperature is constant. This process effectively links the temperature evolution, controlled by $c_s$, to the final spectra evaluation.

%%%%%%%%%%%%%%%%%%%%%%%%%%%%%%%%%%%%%%%%

Further, it is informative to calculate the average transverse velocity of the fireball in terms of the parameters on the freezeout hypersurface. The average transverse velocity can be defined as
\begin{equation}\label{eqn:cyl_avr_def}
    \langle v_T \rangle = \frac{\int v_\rho \sqrt{d\Sigma_\mu d\Sigma^\mu}}{\int \sqrt{d\Sigma_\mu d\Sigma^\mu}}.
\end{equation}
With $v_\rho=\rho/t$ and employing Eqs.~\eqref{eqn:cyl_hyp} and \eqref{eqn:cyl_fo_cond}, we obtain the final expression for the average transverse velocity as
\begin{equation}\label{eqn:cyl_avr}
    \langle v_T \rangle = \sqrt{1+\nu^2}-\nu^2 \log\left[\frac{\nu}{\sqrt{1+\nu^2}-1}\right].
\end{equation}
It is again straightforward to verify that the maximum value of $\langle v_T \rangle$ is $1$ for $\nu=0$. 

It is important to understand the physical meaning of the freezeout parameter $\nu$. To achieve this, we have calculated the average radial and transverse velocities in Eqs.~\eqref{eqn:sph_avr} and \eqref{eqn:cyl_avr}. In Fig.~\ref{fig:avgv}, we show the $\nu$-dependence of average radial velocity for spherical expansion, given in Eq.~\eqref{eqn:sph_avr}, as well as average transverse velocity for cylindrical expansion, given in Eq.~\eqref{eqn:cyl_avr}. We find a monotonic decrease of the average velocities as a function of $\nu$; recall that $\nu=\tau_{3f}/R$ has a dimension of inverse velocity. Therefore the observed monotonic decrease is expected. Moreover, we see that for the same value of $\nu$, the average radial velocity is slightly larger than average transverse velocity. In the next section, we shall use the expressions for the hadron momentum distribution, Eqs.~\eqref{eqn_pT_sph} and \eqref{eqn_pT_cyl}, to fit the experimental data for the transverse momentum spectrum of hadrons. The fit parameters, i.e., temperature $T$ and $\nu$ which represent average expansion velocity, would be interpreted in the context of experimental results. Furthermore, in the blast wave model, the parameter $\beta$ is treated as a fitting parameter, whereas in our approach the average transverse velocity is directly obtained from the hydrodynamic profile at freeze-out.

%%%%%%%%%%%%%%%%%%%%%%%%%%%%%%%%%%%%%%%%%%%%%%%%%%%%%%

\section{Numerical results and discussions}
\label{sec:results}

%------------------------
\begin{figure}[t]
\includegraphics[width=8.6cm]{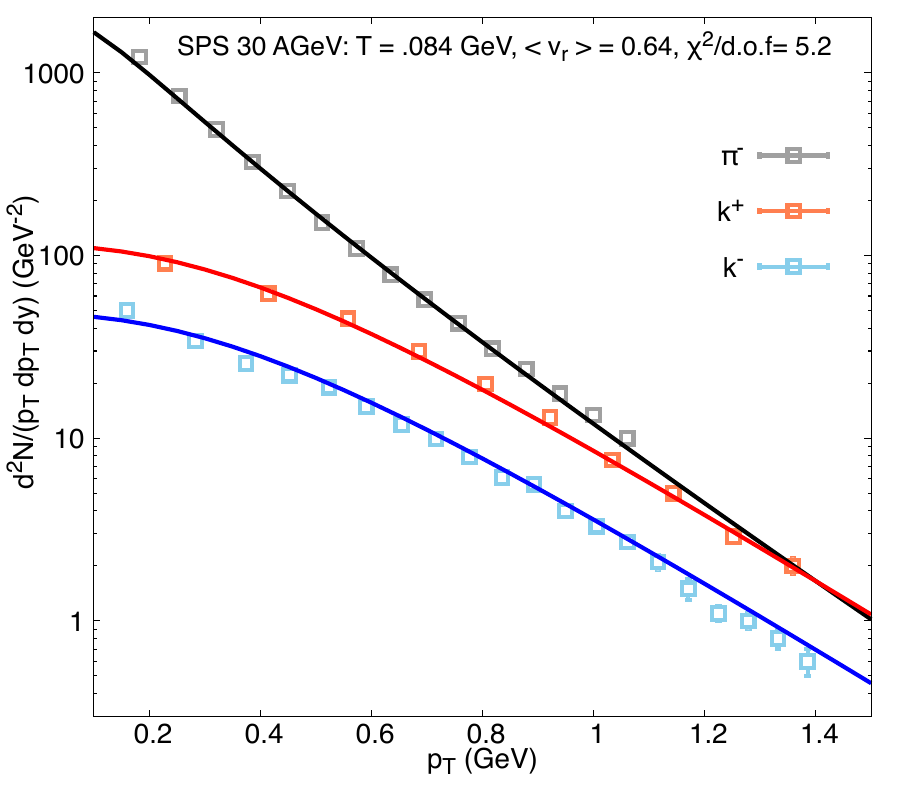}
\includegraphics[width=8.6cm]{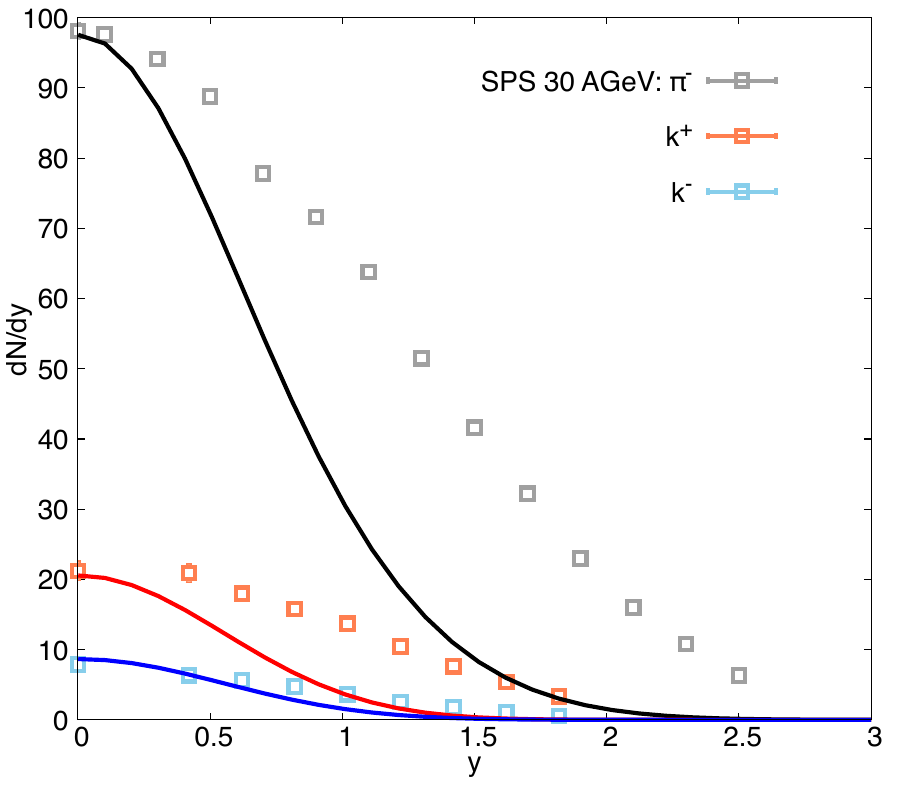}
\caption{ (Left) mid-rapidity $p_T$ spectra for $\pi^-$(black), $k^+$ (red) and $k^-$ (blue) at SPS collision energy of 30 AGeV. Square boxes denote the data from Ref.~{\cite{NA49:2007stj}} and the line represents the model estimations. (Right) $p_T$ integrated rapidity spectra $dN/dy$ of the same species of particles calculated using parameters from fit to $p_T$ spectra. }
\label{fig:resBES}
\end{figure}
%------------------------

In this section, we apply our exact analytical results for hadron momentum spectra to fit experimental data. In low-energy heavy-ion collisions, one expects the expansion geometry of the fireball to be spherical. To this end, we apply our analytical results for spherical geometry to Pb-Pb collisions at $30$~AGeV in Super Proton Synchrotron (SPS). We consider $y=0$ in Eq.~\eqref{eqn_pT_sph} to obtain the transverse momentum, $p_T$, distributions corresponding to the mid-rapidity range. We then employ chi-square minimization to obtain the best-fit result by analyzing the SPS experimental data for mid-rapidity $p_T$ spectra. In this case, apart from the overall normalization, there are two fit parameters, $T$ and $\nu$ (or equivalently $\langle v_r \rangle$). In the left panel of Fig.~\ref{fig:resBES}, we have successfully characterized the charged pion and kaon spectra using our calculated spectra, with a freeze-out temperature of $81$ MeV and average radial velocity $\langle v_r \rangle = 0.65$. These fitted results are in agreement with earlier analyses~\cite{Rode:2018hlj, Rode:2023vjl}. 

With a suitable fitting of the $p_T$ spectra with a reduced $\chi^2$ of $5.6$, it would be interesting to check the momentum-integrated rapidity spectra. The deviation from boost-invariance necessitates that the rapidity spectra exhibit a Gaussian distribution, initially proposed by Landau \cite{Landau:1953gs}. Recently, in Ref.~\cite{Biswas:2019wtp}, a more generalized form of rapidity spectra has been proposed, accommodating a non-conformal equation of state. The distribution in Eq.~\eqref{eqn_pT_sph} depends on transverse momentum  ($p_T$) and rapidity ($y$) via $a=p/(T \nu)$, where momentum $p=\sqrt{p_T^2 + m_T^2 \sinh(y)^2}$. Integrating the distribution over $p_T$ leads to the rapidity spectrum $dN/dy$. On the other hand, we note that the distribution in Eq.~\eqref{eqn_pT_cyl} is independent of $y$ and depends only on $p_T$. 
%By integrating the transverse momentum distribution $\frac{dN}{p_T dp_T dy}$ from Eq.\ref{eqn_pT_sph}, we can evaluate the $dN/dy$. 
In the present work, we evaluate the rapidity spectra for the SPS collision energy 30A GeV using the parameters $T$ and $\nu$ obtained from fitting the slope of the corresponding mid-rapidity $p_T$ spectra using our analytical results. It is to be noted that the overall normalization is different for different particle species considered in this work. 

In the right column of Fig.~\ref{fig:resBES}, we present the rapidity spectra, computed using the parameters derived from the fitting of the mid-rapidity differential $p_T$ data, as discussed earlier. While the Gaussian-like structure is apparent in the resulting rapidity distribution, there exists a noticeable quantitative discrepancy with the data at higher rapidities. This discrepancy can be attributed to the fact that, for fitting the $p_T$ spectra, mid-rapidity data was employed, whereas the same parameter set was utilized for computing this differential rapidity variation across the entire rapidity range. Moreover, the assumption of a completely spherical geometry for this SPS collision energy may not be valid. For a suitable agreement between the data and model estimation of the rapidity spectra, one has to perform fits to the $p_T$ spectra for different rapidity windows, as discussed in Ref.~\cite{Waqas:2021enm}, which is beyond the scope of the present study. 

We note that the feed-down contribution from higher mass resonances are not incorporated here and the fitting is performed considering the primary hadrons only. In the lower momentum region, decay contributions from higher mass resonances dominate total pion yields. At the LHC energy, the fits for pions are customarily performed for $p_T > 0.5$ GeV/c, whereas the $p_T$ range for $k$ and $p$ are $0.2-1.5$ GeV/c and $0.3–3$ GeV/c respectively~\cite{ALICE:2013mez}. For the SPS energy, the pion spectra from Ref.~{\cite{NA49:2007stj}} are weak decay corrected and the fit are generally performed for the interval $0.2 < m_T - m < 0.7$ GeV. To perform state-of-the-art estimation of the freeze-out parameters, one can consider the decay contributions separately following the process mentioned in Ref.~\cite{Rode:2018hlj}. Nevertheless, this work focuses on the analytical solutions and their connections with the spectra. Keeping this in consideration, we perform the simultaneous fits for the available $p_T$ range, considering only the primary hadrons, and have provided qualitative descriptions consistent with earlier works. 

% The experimental data that we consider for the fitting procedure consists of pions with $p_T>0.5$ GeV/$c$ at the LHC energy. The corresponding $p_T$ range for kaons ($K$) and protons ($p$) are $0.2–1.5$ GeV/$c$ and $0.3–3$ GeV/$c$ respectively~\cite{ALICE:2013mez}. For the SPS energy, we use the pion spectra from Ref.~{\cite{NA49:2007stj}} which are weak decay corrected. We perform the fitting in the interval $0.2 < m_T - m < 0.7$ GeV. Since this work focuses on the analytical solutions and their connection with the spectra, we do not consider the contributions from resonance decays for the fitting procedure for simplicity. The freeze-out parameters can be extracted, however, in a state-of-the-art way following the process mentioned in Ref.\cite{Rode:2018hlj}.

%------------------------
\begin{figure}[t]
\includegraphics[width=8.6cm]{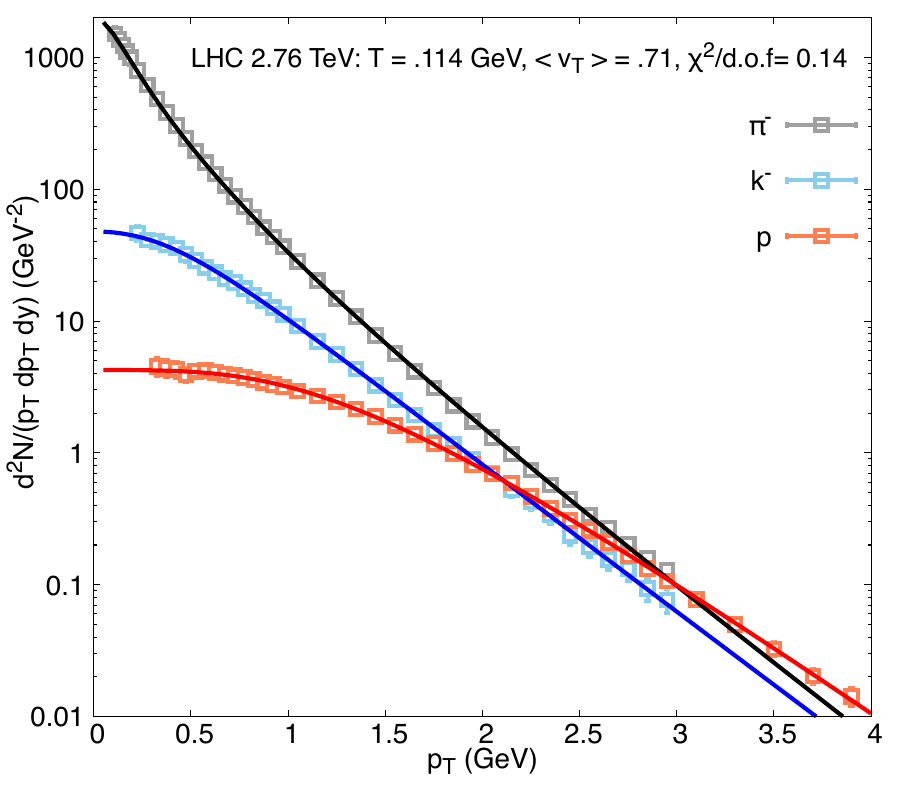}
\caption{(Left) mid-rapidity $p_T$ spectra for $\pi^-$(black), $k^-$ (blue) and $p$ (red) at LHC collision energy of 2.76 ATeV. Square boxes denote the data from Ref.~\cite{ALICE:2013mez} and the line represents the model estimations.}
\label{fig:resLHC}
\end{figure}
%------------------------

At ultra-relativistic high energies, the system exhibits longitudinal boost-invariance, a characteristic we have incorporated into our analytical hydrodynamics framework by adopting cylindrical geometry with boost-invariance along the $z$ direction. This solution has been translated into the form of transverse momentum spectra as given in Eq.~\eqref{eqn_pT_cyl}. We have applied this result to fit the transverse momentum spectra observed in LHC collisions at $\sqrt{s_{NN}}=2.76$~TeV energy. Utilizing mid-rapidity $p_T$ spectra data for pions, kaons, and protons, we conducted a chi-square minimization process. The fitted spectra are plotted along with the data in Fig.~\ref{fig:resLHC}. The resulting temperature is found to be $112$ MeV, which is higher compared to the value observed at SPS ($81$ MeV). These reference values align with our general expectation of a higher freeze-out temperature at LHC due to the increased collision energies. Additionally, we obtain $\langle v_T\rangle = 0.72$ at LHC which aligns with the conventional blast wave parameterization detailed in Ref.~\cite{ALICE:2013mez}. In contrast, the average radial velocity ($\langle v_r \rangle$) at SPS energies is lower, indicating a higher value of $\nu$. This observation suggests a relatively later freeze-out time at lower collision energies.

%%%%%%%%%%%%%%%%%%%%%%%%%%%%%%%%%%%%%%%%%%%%%%%%%%%%%%

\section{Summary and conclusions}
\label{sec:summary}

Analytical hydrodynamic solutions play a crucial role in establishing a benchmark for understanding the evolution of QCD matter created in heavy-ion collisions. While providing a general solution remains challenging, one can focus on specific geometries and flow profiles that facilitate formulation. In this study, we obtained analytical solutions for relativistic hydrodynamics in a system characterized by cylindrical symmetry with boost-invariant longitudinal expansion and Hubble-like transverse expansion. Additionally, we explored an analytical solution for a Hubble-like spherically expanding system. These straightforward profiles allowed us to derive a simple analytical solution for the temperature evolution relative to the proper time, facilitating a connection with the freeze-out hypersurface. For both cylindrical and spherical flow profiles, we computed the analytical expressions for the transverse momentum spectra of hadrons on a constant temperature freeze-out hypersurface, utilizing the Cooper-Frye prescription. 

At this juncture we would like to note that the results for analytical solutions of relativistic hydrodynamic equations have been obtained earlier \cite{Csorgo:2003rt, Lin:2009kv}. In separate works, the analytical expressions for hadron spectra have also been obtained earlier within a phenomenologically motivated paramterized blast-wave model \cite{Florkowski:2010zz, Heinz:2004qz}. In this work, for the first time, we make the connection between the analytical hydrodynamic solutions and the analytical expressions for the hadron spectra. Moreover, we also show that these results are consistent with constant temperature freeze-out hypersurface (as considered in hydrodynamic simulations) rather than the blast wave assumption of constant time freeze-out hypersurface. Further, we have derived the analytical expressions for average radial and transverse velocity on the freeze-out hypersurface, which are exact results. 

We emphasize that the expressions obtained for hadron momentum distributions are exact analytical hydrodynamic results. Our analytical results can serve to test numerical calculations of particle spectra obtained by solving more realistic hydrodynamical equations. We compared our results for transverse momentum spectra with experimental data from SPS and LHC energies. Adhering to the symmetry of the expanding system, we have employed the spherical solution for lower-energy collisions at SPS while opting for the cylindrical description at the LHC. We used $\chi^2$ minimization to analyze available $p_T$ spectra data for SPS collisions at $30$ AGeV and LHC collisions at $2.76$ TeV. The successful description of the data with our analytical results provides a validation to this solution. However, we note that the inclusion of equation of state in the analytical solutions to hydrodynamic equations are rather restrictive and does not allow for incorporating realistic equation of state such as lattice QCD results. On the other hand, it may be possible to extract an average speed-of-sound from analysis of experimental observables. A more detailed phenomenology of heavy-ion collision experiments, within the presented framework, is left for future work.

\section*{Acknowledgments}
D.B. is supported by the Department of Science and Technology, Government of
INDIA under the SERB National Post-Doctoral Fellowship Reference no. PDF/2023/001762.

\bibliographystyle{elsarticle-num}
\bibliography{transverse}
\end{document}